\documentclass[]{iau}

\usepackage{amsmath}
\usepackage{graphicx}
\usepackage{multirow}
\usepackage{wrapfig}
\usepackage{longtable}
\usepackage{comment}

\begin{document}
\newcommand\arcdeg{\mbox{$^\circ$}}
\newcommand\arcmin{\mbox{$^\prime$}}
\newcommand\arcsec{\mbox{$^{\prime\prime}$}}
\newcommand{\farcs}{\hbox{$.\!\!^{\prime\prime}$}} 
\newcommand{\kms}{km~s$^{-1}$}
\newcommand{\wm}{W~m$^{-2}$}
\newcommand{\wmhz}{W~m$^{-2}$~Hz$^{-1}$}
\newcommand{\micron} {$\mu$m}
\newcommand{\s}{\mbox{$''$}}
\newcommand{\mloss}{\mbox{$\dot{M}$}}
\newcommand{\my}{\mbox{$M_{\odot}$~yr$^{-1}$}}
\newcommand{\lsun}{\mbox{$L_{\odot}$}}
\newcommand{\ms}{\mbox{$M_{\odot}$}}
\newcommand{\rs}{\mbox{$R_{\odot}$}}
\newcommand{\mdot}{$\dot{M}  $}
\newcommand{\sect}[1]{\S\ref{#1}\xspace}
\newcommand{\cplus}{C$^{+}$}
\newcommand{\nplus}{N$^{+}$}

\newcommand{\vexp}{\mbox{$V_{\rm exp}$}}
\newcommand{\vsys}{\mbox{$V_{\rm sys}$}}
\newcommand{\vlsr}{\mbox{$V_{\rm LSR}$}}
\newcommand{\tex}{\mbox{$T_{\rm ex}$}}
\newcommand{\teff}{\mbox{$T_{\rm eff}$}}
\newcommand{\tmb}{\mbox{$T_{\rm mb}$}}
\newcommand{\trot}{\mbox{$T_{\rm rot}$}}
\newcommand{\tkin}{\mbox{$T_{\rm kin}$}}
\newcommand{\dens}{\mbox{$n_{\rm H_2}$}}
\newcommand{\bri}{\mbox{erg\,s$^{-1}$\,cm$^{-2}$\,\AA$^{-1}$\,arcsec$^{-2}$}}
\newcommand{\brib}{\mbox{erg\,s$^{-1}$\,cm$^{-2}$\,arcsec$^{-2}$}}
\newcommand{\flux}{\mbox{erg\,s$^{-1}$\,cm$^{-2}$\,\AA$^{-1}$}}
\newcommand{\ha}{\mbox{H$\alpha$}}

\newcommand{\aua}{\textit{A$\,$\&$\,$A}}            
\newcommand{\araa}{\textit{ARA$\,$\&$\,$A}}          
\newcommand{\aap}{\textit{A$\,$\&$\,$A}}
\newcommand{\aapr}{\textit{A\,$\&$\,ARv}}
\newcommand{\aar}{\textit{A$\,$\&$\,$AR}}           
\newcommand{\aas}{\textit{A$\,$\&$\,$AS}}           
\newcommand{\apj}{\textit{ApJ}}                     
\newcommand{\apjs}{\textit{ApJS}}                   
\newcommand{\apjl}{\textit{ApJ Let}}                
\newcommand{\aj}{\textit{AJ}}                       
\newcommand{\sva}{\textit{SvA}}                     
\newcommand{\pasp}{\textit{PASP}}                   
\newcommand{\pasa}{\textit{Publ. Astr..\,Soc.\,of\,Australia}}                   
\newcommand{\pasj}{\textit{PASJ}}                   
\newcommand{\mnras}{\textit{MNRAS}}                 
\newcommand{\nat}{\textit{Nature}}
\newcommand{\baas}{\textit{BAAS}}                   
\newcommand{\aass}{\textit{A$\,$\&$\,$Spa. Sci.}}   
\newcommand{\spscirev}{\textit{Space Sci. Rev.}}

\lefttitle{Sahai \& PANORAMA consortium}
\righttitle{The PANORAMA survey}

\journaltitle{Planetary Nebulae: a Universal Toolbox in the Era of Precision Astrophysics}

\jnlDoiYr{2023}
\doival{10.1017/xxxxx}
\volno{384}

\aopheadtitle{Proceedings IAU Symposium}
\editors{O. De Marco, A. Zijlstra, R. Szczerba, eds.}

\title{High-Speed Outflows and Dusty Disks during the AGB to PN Transition: The PANORAMA survey}

\author   
{Raghvendra Sahai$^1$,
Javier Alcolea$^{2}$,
Bruce Balick$^{3}$,
Eric G. Blackman$^{4}$,
Valentin Bujarrabal$^{2}$,
Arancha Castro-Carrizo$^{5}$,
Orsola De Marco$^{6}$,
Joel Kastner$^{7}$,
Hyosun Kim$^{8}$,
Eric Lagadec$^{9}$,
Chin-Fei Lee$^{10}$,
Laurence Sabin$^{11}$,
M. Santander-Garcia$^{2}$,
Carmen S\'anchez Contreras$^{12}$,
Daniel Tafoya$^{13}$,
Toshiya Ueta$^{14}$,
Wouter Vlemmings$^{13}$,
Albert Zijlstra$^{15}$
}

\affiliation{$^1$ Jet Propulsion Laboratory, Pasadena, CA, USA, 
$^{2}$ Observatorio Astronómico Nacional (IGN/CNIG), MITMA, Spain,
$^{3}$ Dept. of Astronomy, University of Washington, Seattle, WA 98195-1580, USA, 
$^{4}$ Dept. of Physics \& Astronomy, U. Rochester \\ Rochester, NY, 14621, USA, 
$^{5}$ Institut de Radioastronomie Millimétrique, 300 rue de la Piscine, 38406 Saint-Martin-d'Heres, France,
$^{6}$ School of Mathematical and Physical Sciences, Macquarie University, Sydney, New South Wales, Australia,
$^{7}$ Rochester Institute of Technology, Rochester, NY 14623 USA,
$^{8}$  Korea Astronomy and Space Science Institute, 776, Daedeokdae-ro, Yuseong-gu, Daejeon 34055, Republic of Korea,
$^{9}$ Université Côte d'Azur, CNRS, Boulevard de l'observatoire, CEDEX 4, 06304 Nice, France,
$^{10}$ Academia Sinica Institute of Astronomy and Astrophysics, No. 1, Sec. 4, Roosevelt Road, Taipei 106, Taiwan, 
$^{11}$ Instituto de Astronom{\'i}a, Universidad Nacional Aut\'onoma de M\'exico, Apdo. Postal 877, C.P. 22860, Ensenada, B.C., M\'exico, 
$^{12}$ Centro de Astrobiolog{\'i}a (CAB), CSIC-INTA. Postal address: ESAC, Camino Bajo del Castillo s/n, E-28692, Villanueva de la Ca\~nada, Madrid, Spain,
$^{13}$ Chalmers University of Technology, Onsala Space Observatory, Onsala, Sweden,
$^{14}$ University of Denver, USA,
$^{15}$ Jodrell Bank Centre for Astrophysics, Dept. of Physics and Astronomy, University of Manchester, Manchester, UK
}

\begin{abstract}
As mass-losing asymptotic giant branch (AGB) stars evolve to planetary nebulae (PNe), the mass outflow geometries transform from nearly spherical to extreme aspherical. The physical mechanisms governing this transformation are widely believed to be linked to binarity and the associated production of disks and fast jets during transitional (post-AGB) evolutionary stages. We are carrying out a systematic ALMA survey ($P$re-planet$A$ry $N$ebulae high-angular-res$O$lution su$R$vey with $A$L$MA$ or PANORAMA) of a representative sample of bipolar and multipolar post-AGB objects. We have obtained high angular-resolution ($0\farcs1-0\farcs4$) observations of the CO(3--2) and/or 6--5 emission in order to probe the spatio-kinematic structure of the collimated outflows and the central disk/tori. The results are remarkable, generally showing the presence of bipolar or multipolar high-velocity outflows, dense toroidal waists, and in one case, a geometrically-thin circular ring around the central bipolar nebula. A high degree of point-symmetry characterises the morphology of the mass ejecta. In this contribution, we present these and other highlights from our survey. We aim to use 2D/3D radiative transfer modeling in order to derive accurate outflow momenta, masses and mass-loss rates for our sample, and build hydrodynamical models that can explain the observed spatio-kinematic structures. These results will then be used to distinguish between different classes of PN-shaping binary interaction models.
\end{abstract}

\begin{keywords}
Surveys, stars: binaries, planetary nebulae
\end{keywords}

\maketitle

\section{Introduction}
Understanding the impact of binary interactions on stellar evolution is a major challenge -- these interactions dominate a substantial fraction of stellar phenomenology \citep{ivanova, bond, tylenda}, and likely play a major role in the formation of most Planetary Nebulae (PNe), objects that represent the bright end-stages of most stars in the Universe. Such interactions can help explain why, even though the progenitors of PNe -- AGB stars and their circumstellar envelopes -- are generally slowly-expanding ($V_{exp}\sim$\,5-15\,\kms) round structures, the vast majority of PNe deviate strongly from spherical symmetry, showing a dazzling variety of bipolar, multipolar, and elliptical morphologies  \citep{st98,ueta00,setal11pnhst,stanghellini16}. A morphological survey of objects in the evolutionary transition stage between AGB stars and PNe, i.e., pre-planetary nebulae (PPNe), shows a complete lack of round objects (\cite{setal07ppnhst}, hereafter $Setal07$; \cite{siodmak}; \cite{lagadec11}). In addition, PPNe also show the presence of fast outflows ($V_{exp}\gsim$\,50-100\,\kms), which, when resolved (using interferometric observations), are generally highly-collimated. Dense, dusty, equatorial waists are also frequently found in PPNe and PNe, and recognized as an important morphological feature of this class of objects ($Setal07$, \cite{setal11pnhst}). About half of post-AGB objects that are known (or likely) binaries show prominent disks \citep{winckel03} (dubbed ``dpAGB" objects: \citealt{setal11vla}), and have very weak (or no) outflows. 

The current consensus is that the primary PN formation/shaping process is hydrodynamic sculpting of the progenitor AGB mass-loss envelopes, from the inside-out by wandering and/or episodic jets during the short-lived ($\sim1000$\,yr) PPN phase \citep{st98,soker02,balfrk02}. Support for this conclusion is provided by the fact that the momenta associated with PPN outflows, as inferred from analysis of CO observations of PPNe, either single-dish (e.g., \cite{bujarrabal01}) or interferometric (mostly with angular resolution poorer than $\sim1{''}-2{''}$ (e.g., \cite{cox2688,alcolea07,arancha10,sah08005,hans19}) show that these cannot be radiatively-driven. The jets are believed to result from the binary interaction; magnetic fields may play a role in PN-shaping as well, but are also most likely induced by close binaries (e.g., \cite{garciasegura16}).

Hydrodynamic simulations of close binary interaction (e.g., leading to common-envelope evolution) have long struggled to describe even the simplest cases (e.g., \cite{reichardt19}), hence quantitative models that can explain the resulting jet and waist formation are lacking. An analytical approach by \cite{bl14} (hereafter $BL14$) predicts specific observational signatures corresponding to various possible modes of binary interaction. $BL14$ used the (limited) outflow data known for a sample of PPNe at ``face-value", to derive the minimum required mass-accretion rates (\mdot$_a\,\propto\,M_j\,V_j/t_{acc}$), where $M_j\,V_j$ is the total jet momentum and $t_{acc}$ is the accretion time-scale, and inferred that accretion modes such as Bondi-Hoyle-Lyttleton (BHL) wind-accretion and wind Roche lobe overflow (wRLOF) were too weak to power these outflows. However, successful application of BL14's method needs an accurate determination of the fast outflows' physical properties in a representative morphological sample of PPNe, especially $M_j\,V_j$, and $t_{acc}$. 

To this end, we are carrying out a systematic program of high-fidelity, $0\farcs1$ ALMA imaging of PPNe ($P$re-planet$A$ry $N$ebulae high-angular-res$O$lution su$R$vey with $A$L$MA$ or PANORAMA). From our observations, we will estimate the minimum mass-accretion rate using (a) \mdot$_a= Q\,\sum_{k=1}^{n} M_{j,obs}\,V_{j,obs} / t_{accr(k)} = \sum_{k=1}^{n} \int\,dm(V_{off})\times|V_{off}|\,/ (sin(i_k)\,t_{accr(k)})$ (using eqns. 2 from $BL14$ and \cite{bujarrabal01}), where $dm(V_{off})$ is the emitting mass in a small velocity interval at an offset velocity $V_{off}=V_{lsr}-V_{sys}$, $n$ is the number of multiple outflow-pairs, $i_k$ is the inclination angle relative to the sky-plane and $t_{accr(k)} \lsim\,t_{exp(k)}$ is the acretion time-scale for the $k$'th outflow-pair, and $Q$ is a numerical factor typically satisfying $1<Q<5$ in jet models. For a bipolar outflow, n=1. The value of $dm(V_{off})$ is sensitive to the excitation temperature, $T_{ex}(V_{off})$, which we will estimate using the CO(6--5)/CO(3--2) ratio as a function of $V_{off}$.

\section{The PANORAMA Survey}
The PANORAMA survey covers a small (8), but representative sample of bipolar and multipolar Pre-Planetary Nebulae (PPNe), with the ultimate goal of providing quantitative constraints on the shaping mechanism(s). All objects have been previously imaged in the optical with HST, allowing a detailed classification of their morphology ($Setal07$). 
The survey sample includes objects that are (i) at the high end of the range of observed fast-outflow momenta ($M_j\sim 10^{39}$\,g\,cm\,s$^{-1}$: extreme outflow object IRAS\,19374) or at the low end ($M_j\sim 10^{36}$\,g\,cm\,s$^{-1}$: IRAS\,19024, \cite{sah19024}); (ii) with different kinds of point-symmetry, indicative of different kinds of jet activity -- e.g., $PS(s)$: continuous jet with time-variable axis as in IRAS\,10197; $PS(an)$: episodic jet with time-variable axis, or simultaneous multiple jets as in IRAS\,19024; and (iii) with and without halos (e.g., IRAS\,19374 \& IRAS\,17106), i.e., with high and low progenitor AGB mass-loss rates, respectively.\vspace{0.1in}

\begin{table}
\caption{Target Sample: Physical Properties}\label{tbl:targets}
\begin{tabular}{p{1.7cm} p{0.9cm} p{1cm} p{1cm} p{1.4cm} p{0.6cm} l l}
\hline
Name       & Spec. & $^a$HST & $^b$Point  & $^c$Size  &  Dist. &  $^d$CO Morph.,   &  $^e$Cont-     \\
(IRAS)     & Typ.  & Prim.   & Symm.      & $({''})\times({''})$ & (kpc) &  Kinematics;  &  inuum     \\
           &       & Morph.  & $PS$       &        &       &  (FWZI[\kms]) &  (TM1)     \\
\hline
06530$-$0213 & F\,5Ia &  M  &   s           &  $0.6\times1.1$ & 1.3   &  Mp\,Ofl,W,oR;\,(30)&  NoCen \\ 
08005$-$2356 & F\,5Ie &  B  &   s           &  $0.6\times1.4$ & 3   &  Mp\,Ofl;\,(270)   &  Cen  \\ 
09371+1212   & K\,7III & M & m,\,an,\,s     &  ? & 3     & ?;\,(55)         &  Cen \\ 
10197$-$5750 & A\,2Iab & B &   s            &  $2.1\times3.6$ & 2     &   Bp\,Ofl,W;\,(100)  &  Cen \\ 
17047$-$5650 & WC\,10 &  M  &   m           &  $2.6\times6.2$ & 1.35  &  Mp, W;\,(142)    &  Cen \\ 
17106$-$3046 & F\,5I  &  B   &  s           &  $0.7\times1.1$ & 4     &  Bp\,Ofl;\,(55)    &  Cen($1{''}\times0\farcs45$) \\ 
19024+0044  & G\,0-5  & M &  m,\,an        &  $1.5\times2.1$ & 3.5   &    Mp\,Ofl;\,(100)   &  NoCen \\ 
19374+2359  & B\,3-6  & B  &  s            &  $1.8\times2.4$ & 5     &   Mp\,Ofl;\,(375)   &  NoCen \\ 
\hline
\end{tabular}
\noindent $Notes\,to\,Table$: (a) Optical (HST) Prim. Morph.: B=Bipolar, M=Multipolar; (b) Point Symmetry: multiple pairs of lobes=$PS(m)$, ansae=$PS(an)$, shape=$PS(s)$ (details in \cite{setal11pnhst}); (c) Semi-minor and semi-minor axes of smallest ellipse that includes observed CO emission (for IRAS\,06530, the size refers to the inner aspherical nebula, and excludes the outer ring); (d) CO (ALMA) Prim. Morph. (from TM1 \& TM2 data): Bp\,Ofl=bipolar outflow, Mp\,Ofl=Multipolar Outflows, W=waist/torus, oR=outer ring, ps=point-symmetry present, ?=unclear (most flux apparently resolved out), FWZI=full-width at zero-intensity; (e) Continuum (ALMA): NoCen=no centrally-peaked continuum source in TM1 observations, Cen=centrally-peaked continuum source [FWHM (major$\times$minor) if resolved].\\
\end{table}

\begin{figure}[h]
\begin{center}
 \includegraphics[width=13.2cm]{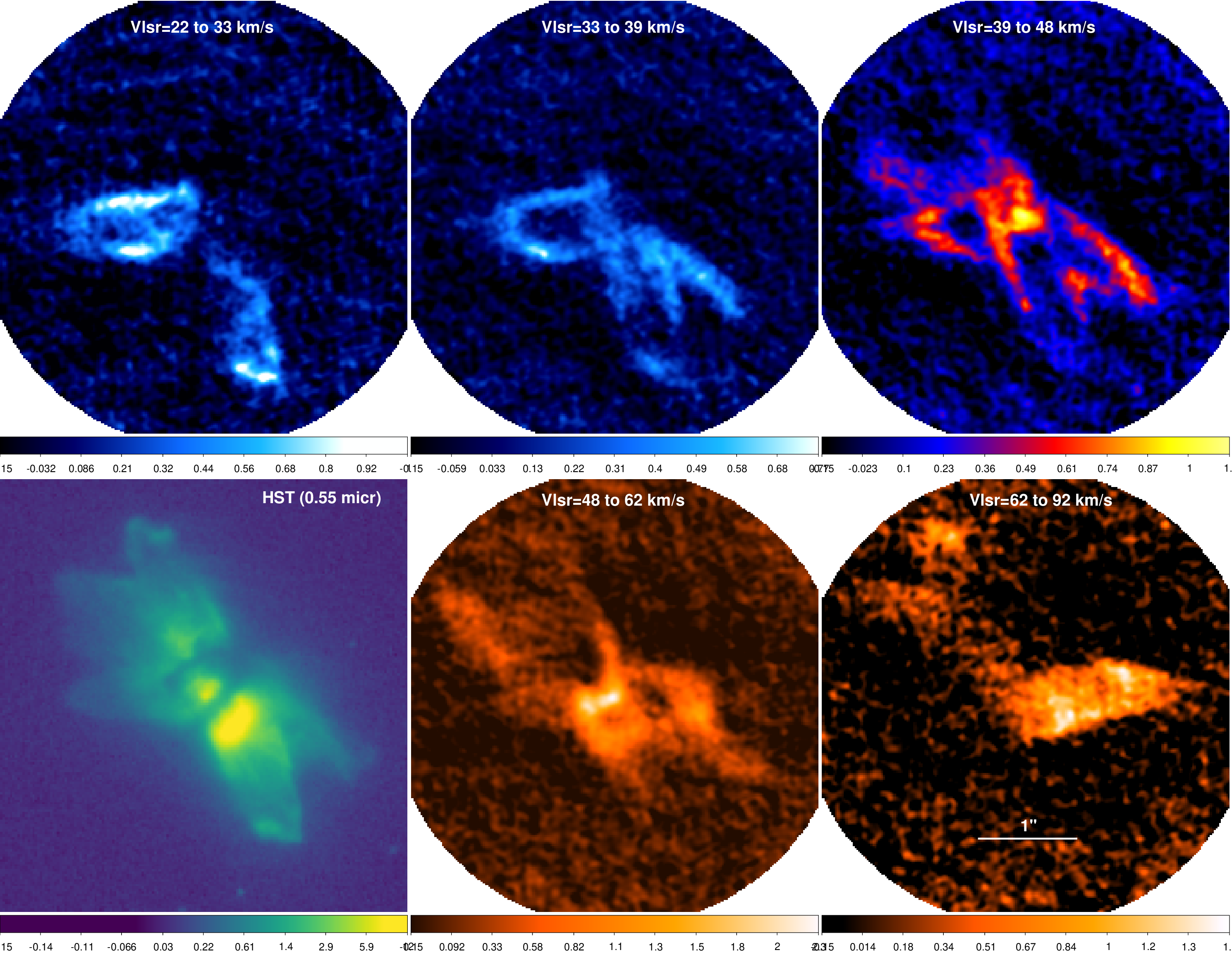}
\end{center}
\vspace{-0.2in}
\caption{CO(6--5) TM1 images of IRAS\,19024 in different velocity ranges within its spectrum chosen to show the shapes of its multipolar lobes for comparison with those seen in scattered light with HST in the optical. The systemic velocity is $V_{lsr}\sim50$\,\kms. 
}
\label{i19024-co65} 
\end{figure}

\begin{figure}
\begin{center}
 \includegraphics[width=13.2cm]{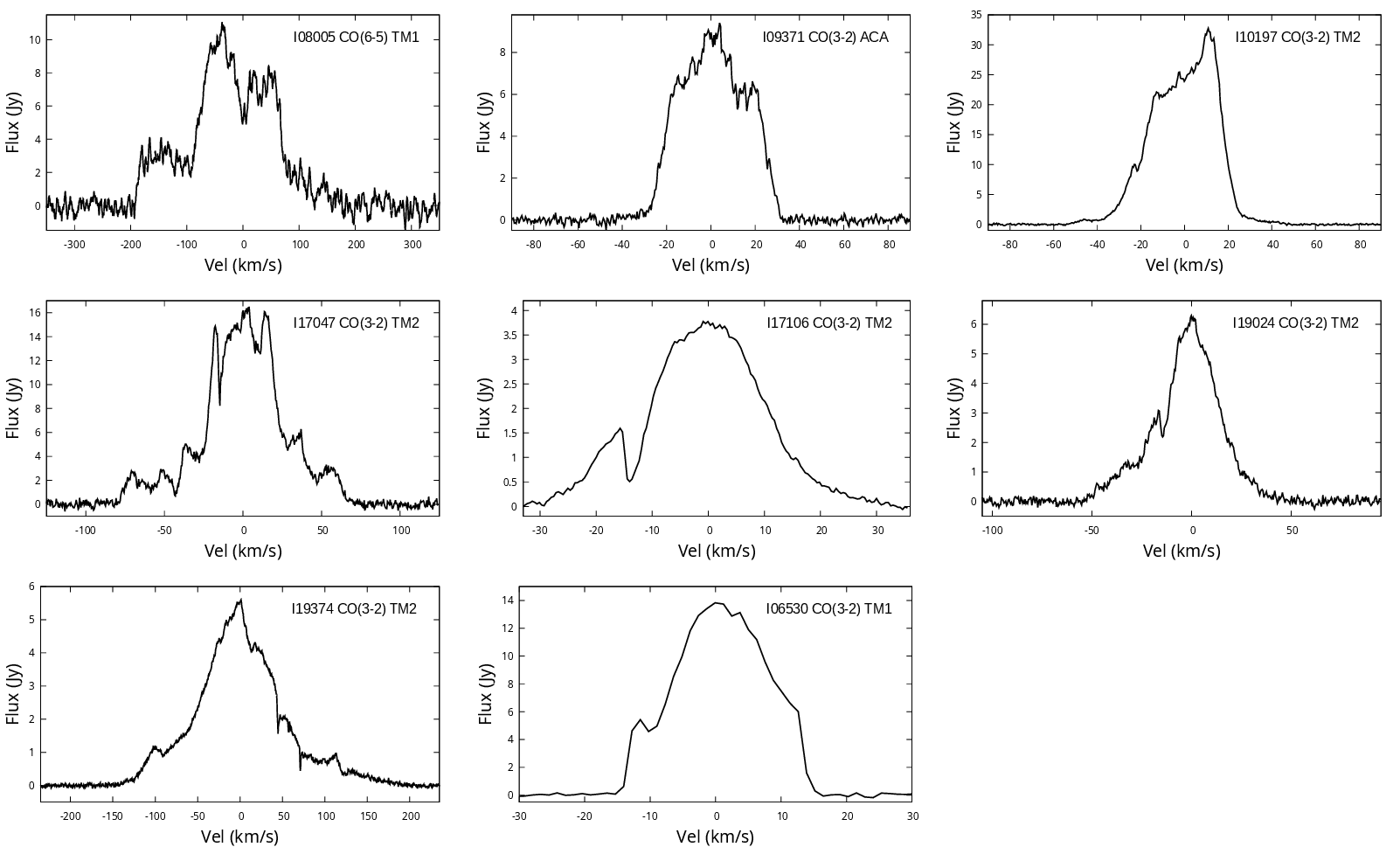}
\end{center}
\vspace{-0.2in}
\caption{Representative (spatially-integrated) CO J=6-5 or 3-2 spectra, for the PANORAMA sample of PPNe. The systemic velocity has been set to zero for each object.
}
\label{ppnspec} 
\end{figure}

\noindent\underline{Survey\,Highlights}: Although the survey, requiring observations with the 12-m array in the TM1, TM2 configurations and the 7-m array (with angular resolutions of $\sim0\farcs1$, $\sim0\farcs5$, $\sim5{''}$, respectively) is still in progress, the existing data includes TM1 imaging of all 8 objects in one or both of Bands 7 and 9, which include the CO(3--2) and CO(6--5) lines, respectively. The results are remarkable, generally showing the presence of extended, collimated lobe-structures in CO that presumably result from bipolar or multipolar ($V_{exp}\sim\,30$\,\kms) outflows (Figs.\,\ref{i19024-co65},\ref{ppnspec}). For IRAS\,09371, the imaging suffers from strong artifacts that obscure its structure \citep{arancha05}. The maximum (projected) outflow velocities, estimated from the spectra, can be broadly classified into 3 classes: extreme (i.e., $V_{exp}\gsim70$\,\kms: IRAS\,17047 and IRAS\,19374), medium ($70>V_{exp}>30$\,\kms: IRAS\,09371, IRAS\,10197, IRAS\,17106, and IRAS\,19024) and low (IRAS\,06530: we discuss this object in detail below). Three optically-classified bipolar objects show the possible presence of an additional (fainter) outflow with a different orientation than the optical symmetry axis, indicating that these objects may also be intrinsically multipolar.

The high angular-resolution maps enable us to isolate the dense waists from the low-latitude region of the outflows in the target sample and resolve their true structure. One of the surprising results is the conspicuous absence of gas close to the central star in the majority of objects, suggesting that the waist structures have a toroidal structure, resulting from an episode of equatorially-enhanced mass-loss from the progenitor AGB, as for e.g., in a CE system (e.g., \cite{sandq98,pssy12}). This result is in marked contrast to dpAGB objects, such as the Red Rectangle, AC\,Her, and IW\,Car \citep{buj13,buj15,buj17}, that show compact rotating disks of gas+dust, with masses of $\sim0.0012-0.006$\,\ms; these may have formed due to enhanced RLOF ($BL14$). 

\begin{figure}
\begin{center}
 \includegraphics[width=13.2cm]{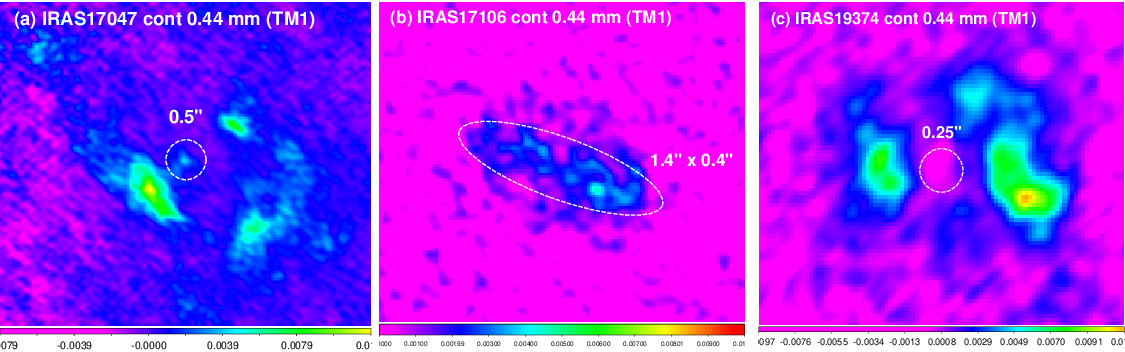}
\end{center}
\vspace{-0.2in}
\caption{Three classes of continuum emission (0.44\,mm TM1 data) (a) extended emission + compact central source, (b) extended central source, (c) extended emission, no central source. The dashed circles or ellipse (with sizes) show the location of the central star.
}
\label{ppn-cont} 
\end{figure}

Continuum emission was detected towards all sources in the existing data, which can be divided into 3 broad classes, using the TM1 images (i) extended emission + unresolved central source, (ii) extended central source, (iii) extended emission and no central source (e.g., Fig.\ref{ppn-cont}). These data emphasize the importance of high-resolution observations to characterize the origin and nature of the (sub)mm emission, since lower-reolution studies, with angular resolution $\gsim0\farcs5$ would tend to merge the compact and extened emission into one structure, which could then be mis-interptreted as being a dpAGB-type disk, for example. \vspace{0.05in}

\begin{figure}
\begin{center}
 \includegraphics[width=13.2cm]{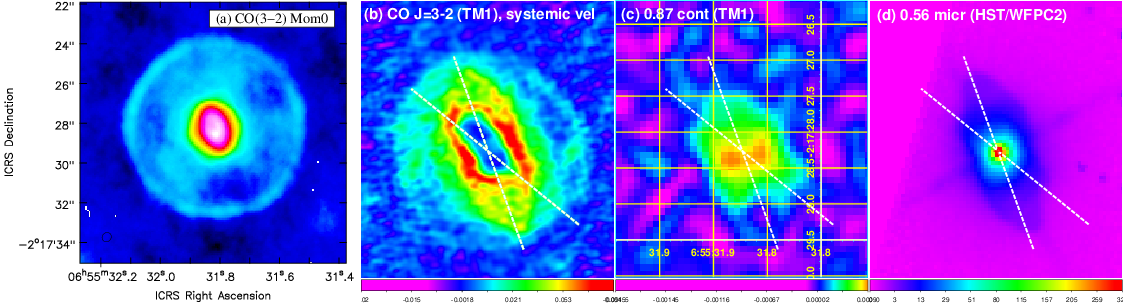}
\end{center}
\vspace{-0.2in}
\caption{The multipolar PPN IRAS\,06530: (a) Moment\,0 map of the CO(3--2) (TM2), (b) Map of CO(3--2) at the systemic velocity showing the structure of the central bipolar nebula (TM1), (c) 0.89\,mm TM1 continuum image, (d) HST/WFPC2 0.56\,\micron~image of scattered light. The dashed lines show the axes of two pairs of lobes seen in the optical. Panels $b-d$ have the same field-of-view.
}
\label{i06530img} 
\end{figure}

\noindent$IRAS\,06530$: IRAS\,06530 shows a large geometrically-thin ring, surrounding a compact central nebula that has a high degree of point-symmetry ({\bf Table\,\ref{tbl:targets}}). Unlike the other PPNe in our sample, no high-velocity emission is seen -- the value of $V_{exp}$ ($\lsim12.5$\,\kms)~is similar to what is typically found for outflows during the AGB phase. Such rings have been seen in several carbon stars and attributed to an episodic increase in the mass-loss rate due to a thermal pulse. Thus, the ring in IRAS\,06530 is due to the last thermal pulse while its central star was still on the AGB. Following the thermal pulse, there was a steep decrease in the mass-loss rate (by a factor $\gsim10$) for $\sim600$\,yr. During this latter period, the mass loss was not spherically symmetric (Fig.\,\ref{i06530img}a). It was then followed by a new episode of enhanced mass loss immediately preceding the start of the CS's post-AGB evolution; the resulting shell shows extended arc-like structures that are likely part of a 3-D spiral structure resulting from interaction with a companion (such spiral structures are being found in a growing list of AGB stars, see e.g. \cite{decin20}). From the expansion ages of the ring ($5500$\,yr) and waist ($750$\,yr), we infer that IRAS\,06530 transitioned from an AGB star to a post-AGB star $4750$\,yr ago. A full description and analysis of these data will be presented in a forthcoming paper (\cite{sahai24}).

In summary, the PANORAMA survey has produced several exciting results and some unexpected ones as well. The morphology of the molecular gas, in almost all cases shows the presence of collimated outflows that correspond to the lobes seen in optical scattered light, confirming that the latter result from sculpting of the ambient circumstellar envelope (ejected by the prognenitor AGB star). But there is one notable exception, the C-rich PPN IRAS\,06530, which shows an extended ring due to the last thermal pulse. In a few optically-bipolar objects, additional faint outflows misaligned with the symmetry axis are found, indicating that these objects may be multipolar. Compact continuum emission around the central star is seen in some sources, but not in others -- in contrast to dpAGB objects, where such emission is one of their defining characteristics.

\begin{acknowledgements} RS's contribution to the research described here was carried out at the Jet Propulsion Laboratory, California Institute of Technology, under a contract with NASA, and funded in part by NASA via ADAP awards, and multiple HST GO awards from the Space Telescope Science Institute.
\end{acknowledgements}


\end{document}